\documentclass{article}
\usepackage[T1]{fontenc} 
\usepackage[utf8]{inputenc} 
\usepackage{striking,amsmath,cite,url}
\usepackage{graphicx}
\usepackage{color}
\usepackage{booktabs}
\usepackage[bookmarks=false]{hyperref}

\title{Striking a New Chord: Neural Networks in Music Information Dynamics}

\twoauthors
  {Farshad Jafari} {Georgia Institute of Technology \\ {\tt farshad.jafari@gatech.edu}}
  {Claire Arthur} {Georgia Institute of Technology \\\ {\tt claire.arthur@gatech.edu}}

\def\authorname{F. Author, S. Author, and T. Author}

\begin{document}

\maketitle
\begin{abstract}

Initiating a quest to unravel the complexities of musical aesthetics through the lens of information dynamics, our study delves into the realm of musical sequence modeling, drawing a parallel between the sequential structured nature of music and natural language. 

Despite the prevalence of neural network models in MIR, the modeling of symbolic music events as applied to music cognition and music neuroscience has largely relied on statistical models. In this "proof of concept" paper we posit the superiority of neural network models over statistical models for predicting musical events. Specifically, we compare LSTM, Transformer, and GPT models against a widely-used markov model to predict a chord event following a sequence of chords.

Utilizing chord sequences from the McGill Billboard dataset, we trained each model to predict the next chord from a given sequence of chords. We found that neural models significantly outperformed statistical ones in our study. Specifically, the LSTM with attention model led with an accuracy of 0.329, followed by Transformer models at 0.321, GPT at 0.301, and standard LSTM at 0.191. Variable Order Markov and Markov trailed behind with accuracies of 0.277 and 0.140, respectively. Encouraged by these results, we extended our investigation to multidimensional modeling, employing a many-to-one LSTM, LSTM with attention, Transformer, and GPT predictors. These models were trained on both chord and melody lines as two-dimensional data using the CoCoPops Billboard dataset, achieving an accuracy of 0.083, 0.312, 0.271, and 0.120, respectively, in predicting the next chord.

To evaluate their performance comprehensively, we employed various metrics, including cross-entropy loss and Word2Vec similarity, observing a strong correlation between them. To demystify these models' decision-making processes, we conducted a feature attribution and sensitivity analysis, aiming to understand their categorization of musical events. Despite these models' promise, their application to cognition or neuroscience is somewhat masked by the inherent 'interpretability' problem that persists in neural network modeling. This work aims to highlight the utility of deep learning models in music for interdisciplinary research, with the aim of shedding light on cognitive processes behind human musical expectancy, with implications for research in cognitive and neuroscience research into musical expectancy, emotion, and aesthetics.

\end{abstract}
\section{Introduction}\label{sec:introduction}

Despite the fact that the fields of music information retrieval (MIR) and music perception both rely heavily on musical modeling, each field has done so not only with different approaches, but with different aims.
In MIR, historically, modeling some musical process or feature served the aim of designing a tool (software, model, etc.) that would directly apply this feature or model, or benefit and/or use it towards facilitating a related task (e.g., source separation, chord estimation, melody generation, etc.).
On the other hand, in music perception, computational modeling of music has historically served to elucidate some cognitive or musicological process, and models have commonly been used to predict human judgements or behavior in empirical experiments (e.g., expectancy, aesthetic response, emotional reaction, etc.). 
Because the aims in perceptual research rely heavily on understanding causality, statistical modeling has been the standard in the field.\footnote{There are obvious exceptions to this going all the way back to implementations of neural networks in the 1980s, e.g., Bharucha, 1987 \cite{bharucha_music_1987}.}
In MIR, on the other hand, with the focus on the performance of the output rather than the model interpretation, machine learning (ML) approaches have dominated the field for over a decade.

The present paper aims to demonstrate the use of ML in the task of prediction (as opposed to generation, for example), showcasing not only the superior performance of deep learning models for prediction, but also suggesting some methods for interpreting model behavior, thus making them more accessible for use in music cognition research.
In addition, both ML and statistical methods have typically used different evaluation methods.
In this paper we offer a ``proof of concept'', offering a direct comparison of several ML models and a commonly-used statistical model in the context of chord prediction, and evaluate them across several metrics. 
Our aim is twofold: to showcase the relevance and utility of deep learning approaches for music prediction, and to illustrate the relative performance of several common ML models side-by-side across several metrics taken from ML, linguistics, and statistics.


\subsection{Background}

In the field of music perception and cognition, variational Markov-style models have largely dominated the literature in predicting musical structure. Notably, the Information Dynamics of Music (IDyOM) models analyze the informational content of music to predict listener responses and understand the cognitive processes underlying music perception \cite{pearce_construction_2005}. These models have been widely employed to predict cognitive and neuroscientific measures ranging from hedonic reward \cite{cheung_uncertainty_2019}, musical expectation \cite{gingras_linking_2016}, and aesthetic response \cite{pearce_neuroaesthetics_2016}. 

Another significant computational approach is Bayesian inference, a statistical method that updates probability estimates for a hypothesis as more evidence becomes available \cite{vuust_rhythmic_2014}. Building on this concept, the Dynamic-Regularity Extraction (D-REX) model uses Bayesian inference to predict musical surprisal, with studies showing that D-REX outputs significantly correlate with neural signals \cite{abrams_retrieving_2022}.

Advancements in computational modeling have extended to the use of deep neural networks, with diffusion models being applied to model musical expectancy in symbolic data \cite{masclef_deep_2023}. Despite these advancements, there is a notable lack of data experiments on other deep models, specifically sequential models, in modeling musical expectancy. A related domain where these models are applied is music generation, where the task often involves generating music sequentially, akin to prediction, which can inform expectancy models.

The journey of computational music generation can be traced back to one of the first pieces generated by a computer program, "Push Button Bertha" in 1956, which used rule-based random sampling \cite{ames_automated_1987}. Historically, statistical models such as Markov Models laid the groundwork, using joint probabilities of element sequences (m-grams) to synthesize new sequences \cite{brooks_experiment_1957}. Extending this approach, Hidden Markov Models (HMMs) introduced latent states to capture deeper structural dependencies within musical compositions, offering a more nuanced understanding of musical sequences \cite{farbood_analysis_2001}. 

The advent of deep learning heralded a significant shift towards more complex neural architectures in Music Information Retrieval (MIR). Recurrent Neural Networks (RNNs), particularly those utilizing Long Short-Term Memory (LSTM) units, have been pivotal in modeling time-dependent data \cite{boulanger-lewandowski_modeling_2012}. These models excel at capturing the temporal dynamics inherent in music, making them particularly suited for sequence prediction tasks in this domain \cite{waite_generating_2016}. Due to their computational efficiency, RNNs can be deployed in parallel to generate polyphonic music \cite{johnson_generating_2017} and even support real-time interactive music generation systems \cite{hadjeres_anticipation-rnn_2020}.

Building on these advancements, recent models have further pushed the boundaries of music generation. Hierarchical VAEs with recurrent decoders adeptly capture long-term musical structures \cite{roberts_hierarchical_2018}, while VQ-VAEs produce coherent raw audio songs over several minutes \cite{dhariwal_jukebox_2020}. CNN-based GANs have enhanced symbolic music generation \cite{yang_midinet_2017}, and multi-genre GANs facilitate diverse style creation within a unified framework \cite{chauhan_multi-genre_2023}. Additionally, multi-track sequential GANs have improved symbolic music generation and accompaniment, enabling the creation of intricate compositions \cite{dong_musegan_2018}.

Recent advancements have seen Transformer models significantly enhancing music generation by capturing long-term structures \cite{huang_music_2018}, recognizing metrical patterns \cite{huang_pop_2020}, and improving track-level control \cite{ens_mmm_2020}. Multi-track Transformers facilitate intricate multi-instrument compositions \cite{dong_multitrack_2023}, while emotion-conditioned models reflect specific emotional states in generated music \cite{ji_emomusictv_2024}. Transformer-based GANs combine strengths for multi-track generation \cite{zhang_learning_2023}, and pre-trained Transformers fine-tuned on specific datasets tailor music for distinct purposes \cite{donahue_lakhnes_2019}. Lastly, utilizing general-purpose unsupervised technology akin to GPT-2, these models predict the next token in a sequence, whether audio or text, enhancing their versatility \cite{payne_musenet_2019}.

\section{Method}
\subsection{Model Design}

\subsubsection{Markov Model}

The first-order Markov Model applied in our study posits that the probability of transitioning to a subsequent chord is conditional only on the current chord. This assumption is encapsulated in the transition matrix \( P \), where each element \( P_{ij} \) denotes the probability of moving from chord \( i \) to chord \( j \). The transition probabilities are calculated as:
\begin{equation}
P_{ij} = \frac{N_{ij}}{\sum_{k} N_{ik}},
\end{equation}
where \( N_{ij} \) is the number of transitions from chord \( i \) to chord \( j \), and \( \sum_{k} N_{ik} \) is the total number of transitions from chord \( i \) to any chord. A smoothing parameter is introduced to ensure all potential transitions have a nonzero probability, enhancing the model's robustness.

\subsubsection{Variable Order Markov Model}

The Variational Order Markov Model extends the first-order Markov Model by considering varying lengths of chord sequences (contexts) up to a predefined maximum, enabling a more fine-grained prediction of a chord, based on the varied contexts encountered leading up to that chord in the training data. Transition probabilities \( P_{ij|C} \) for moving from context \( C \) (a sequence of chords) to chord \( j \) after chord \( i \) are calculated by:
\begin{equation}
P_{ij|C} = \frac{N_{ij|C} + \alpha}{\sum_{k} (N_{ik|C} + \alpha)},
\end{equation}
where \( N_{ij|C} \) represents the frequency of observing chord \( j \) after context \( C \) and chord \( i \), and \( \alpha \) is a smoothing parameter to handle unseen transitions.

\subsubsection{LSTM Model}

The LSTM model (short for Long Short-Term Memory), is a recurrent neural network model that is particularly suited to sequential data. It addresses the challenge of long-term dependencies in chord sequences by leveraging memory cells and gating mechanisms that allow for optimization and control over useful versus redundant data over long sequences. LSTM models manage sequence data effectively by parameters such as embedding dimension, hidden layer dimension, and the number of layers, which are fine-tuned for optimal prediction accuracy. The core LSTM formula encapsulating its gating mechanics can be represented as follows:
\begin{equation}
f_t = \sigma(W_f \cdot [h_{t-1}, x_t] + b_f),
\end{equation}
where \( f_t \) is the forget gate's activation, \( \sigma \) is the sigmoid function, \( W_f \) and \( b_f \) are the weight and bias for the forget gate, \( h_{t-1} \) is the previous hidden state, and \( x_t \) is the current input vector.

\subsubsection{LSTM with Attention Model}

An LSTM-with-attention model adds another layer of complexity and capability above a generic LSTM model through the inclusion of an attention mechanism that allows for weighted focus on certain parts of the sequence that may be more relevant to the task at hand. In addition, the LSTM-with-Attention is better suited to data containing rather long sequences or complex dependencies. The LSTM-with-Attention model in our implementation calculates a context vector \( c_t \) at each time-step as a weighted sum of the LSTM's hidden states \( h_i \), using computed attention scores:
\begin{equation}
\alpha_t : c_t = \sum_{i} \alpha_{ti} h_i
\end{equation}
The model then combines this context vector with the LSTM's current output through a concatenation operation, followed by a linear transformation to produce the final output \( y_t \). This method allows the model to dynamically focus on relevant portions of the input sequence for improved prediction accuracy of the next chord.

\subsubsection{Transformer Model}

Unlike LSTMs, which process sequences in a linear, sequential order, transformer models abandon the recurrent architecture altogether, and instead use self-attention mechanisms across the entire sequence simultaneously, allowing for the direct connection of distant elements. This architecture is not only more efficient, but affords greater complexity. The Transformer model transforms input sequences \( X \) into embeddings \( X_{\text{emb}} \) through an embedding layer. It employs a Transformer Encoder that utilizes multi-head self-attention and position-wise feedforward networks. The self-attention mechanism is mathematically defined as:
\begin{equation}
    \text{Attention}(Q, K, V) = \text{softmax}\left(\frac{QK^T}{\sqrt{d_k}}\right)V
\end{equation}
The output of this mechanism is processed by a feedforward network, defined as:
\begin{equation}
    \text{FFN}(x) = \text{ReLU}(xW_1 + b_1)W_2 + b_2
\end{equation}
where \( W_1 \), \( W_2 \), \( b_1 \), and \( b_2 \) are network parameters. The sequence's next chord is predicted by applying a linear layer and softmax to the final encoder output, producing the probability distribution over the chord vocabulary:
\begin{equation}
    Y = \text{softmax}(\text{FC}(\text{Encoder}_{\text{output}}))
\end{equation}

\subsubsection{GPT-based Model}

The GPT (Generative Pre-trained Transformer) architecture embodies some of the most advanced capabilities in natural language processing and generation \cite{yenduri_generative_2023}. 
The GPT model, compared to the Transformer model, employs a decoder-like structure with a causal self-attention mechanism for autoregressive sequence generation (i.e., generating one token at a time). This causal attention ensures that each token can only attend to preceding tokens, formulated as:
\begin{equation}
    \text{CausalAtt}(Q, K, V) = \text{softmax}\left(\frac{QK^T}{\sqrt{d_k}} + M\right)V
\end{equation}
where \( M \) is a mask to prevent future tokens from influencing the prediction.

The GPT architecture also modifies the placement of Layer Normalization and integrates it with Position-wise Feedforward Networks within each transformer block, applying normalization before and adding residual connections after each sub-layer. The prediction for the next token in a sequence leverages the accumulated context through:
\begin{equation}
    P(x_{i+1} | x_1, ..., x_i) = \text{softmax}(\mathbf{W}_o h_i)
\end{equation}
These adaptations make the GPT model particularly effective for generative tasks.

\subsubsection{Multi-Feature LSTM Model}

The Multi-Feature LSTM Network is tailored for tasks requiring simultaneous analysis of sequences with multiple features, predicting the target feature's next value by harnessing the collective information from all inputs. Each input feature sequence \( X_{f_i} \) is embedded into a continuous space, represented as \( E_{f_i} = \text{Embedding}(X_{f_i}) \).

Dedicated LSTM layers process each embedded sequence \( E_{f_i} \), yielding latent representations \( H_{f_i} \) for the final timestep, encapsulating each feature's temporal dynamics. These representations are concatenated into a context vector \( C = \text{Concatenate}(H_{f_1}, H_{f_2}, \ldots, H_{f_n}) \), which is then used to predict the next value of the target feature via \( P(y | F) = \text{softmax}(\mathbf{W}_c C + b_c) \), where \( \mathbf{W}_c \) and \( b_c \) are the parameters of the fully connected output layer.

\subsubsection{Multi-Feature LSTM with Attention Model}

The Multi-LSTM with Attention model processes multiple input features in parallel, employing separate LSTM layers for each feature. Attention scores are then aggregated across all features' LSTM outputs, allowing the model to dynamically weigh each feature's importance at every timestep:
\begin{equation}
    \alpha_{f,t} = \frac{\exp(e_{f,t})}{\sum_{f'=1}^{F}\sum_{k=1}^{T} \exp(e_{f',k})}, \quad e_{f,t} = a(h_{f,t}, c_f)
\end{equation}
Here, \( \alpha_{f,t} \) denotes the attention weight for feature \( f \) at timestep \( t \), with \( h_{f,t} \) as the hidden state for feature \( f \) at \( t \), and \( c_f \) as the feature-specific context vector.

\subsubsection{Multi-Feature Transformer Model}

The Multi-Transformer Model processes sequences with multiple features \( F = \{f_1, f_2, \ldots, f_n\} \), using the Transformer architecture to predict the next value of a target feature. 
In our case, we continue to predict the next chord in the sequence, but instead of only considering the prior chord sequence, we also model X, Y, and Z.
Each input feature \( f_i \) is mapped to high-dimensional embeddings \( E_{f_i} \), which are then concatenated to form \( E_{\text{composite}} = \text{Concatenate}(E_{f_1}, E_{f_2}, \ldots, E_{f_n}) \).

The composite embeddings \( E_{\text{composite}} \) are processed by a Transformer, which applies self-attention across all features, represented as \( T_{\text{output}} \). The final prediction \( y \) for the target feature is given by \( P(y | F) = \text{softmax}(\mathbf{W}_t T_{\text{output}} + b_t) \), capturing the interdependencies between features and their influence on the target feature's next value.

\subsubsection{Multi-Feature GPT-based Model}

Unlike the single GPT model, which processes sequences in a uni-dimensional manner, the Multi-GPT Model integrates a feature projection mechanism to align multiple feature sequences into the GPT-2 model's embedding space, enabling multi-dimensional sequence processing. 

In contrast to the Multi-Transformer model, which employs separate embeddings and Transformer blocks for each feature before aggregation, the Multi-GPT Model directly leverages the unified modeling capacity of GPT-2. The adapted GPT-2 architecture processes the projected features, with the last hidden states used for subsequent classification tasks:
\begin{equation}
    \text{Logits} = \text{FC}(\text{GPT2}(\text{Projected\_Features})_{\text{last}})
\end{equation}

\subsection{Evaluation}

\textbf{Accuracy}: Our evaluation framework is centered around assessing the predictive accuracy of the model on unseen test data, specifically measuring its performance in correctly predicting the subsequent chord in a sequence. Accuracy is defined as the proportion of correctly predicted chords to the total number of predictions made.
For harmony (as with language) we understand that despite that the ground truth holds a singular ``correct'' value, in reality other values may substitute equally well.
We discuss the issue of subjectivity in the evaluation in the XXX section. 

\textbf{Perplexity}: In addition to accuracy, we employ perplexity as a secondary metric to evaluate our model's performance. Perplexity measures the model's uncertainty in predicting the next chord, offering insight into its probabilistic forecasting efficacy. Lower perplexity values indicate a higher confidence in predictions and, consequently, a better model performance. Mathematically, perplexity (\(P\)) is defined as the exponential of the average negative log-likelihood of the test set predictions:
\begin{equation}
    P = \exp\left(-\frac{1}{N} \sum_{i=1}^{N} \log(p(x_i))\right)
\end{equation}
where \(N\) is the total number of predictions, and \(p(x_i)\) is the probability assigned to the correct chord \(x_i\) by the model. 
Perplexity is particularly informative in the context of generative models like ours, where the goal extends beyond mere classification to include the generation of probable chord sequences.

\textbf{Word2Vec Similarity}: Beyond accuracy and perplexity, we utilize Word2Vec similarity to evaluate the semantic closeness between predicted and actual chords. This metric is particularly relevant for understanding the model's ability to generate musically coherent predictions, even when not exactly matching the ground truth. The Word2Vec similarity is computed based on the cosine similarity between the vector representations of predicted and actual chords, as obtained from a pre-trained Word2Vec model. The similarity score ranges from 0 to 1, where 1 indicates perfect alignment (or identical vectors) and values closer to 0 denote lower similarity. Formally, the similarity (\(S\)) between two chords \(c_1\) and \(c_2\) is given by:
\begin{equation}
    S(c_1, c_2) = \max\left(1 - \cos(\vec{c_1}, \vec{c_2}), 0\right)
\end{equation}
where \(\vec{c_1}\) and \(\vec{c_2}\) are the Word2Vec embeddings of chords \(c_1\) and \(c_2\), respectively, and \(\cos\) denotes the cosine distance between the two vectors. This metric is essential for cases where the exact chord prediction may not be critical, but the semantic or musical closeness of the prediction to the ground truth holds significance, such as in harmonic and melodic continuity in music composition and analysis.

\subsection{Datasets}

For training and evaluation, we relied on two symbolic music datasets of popular music: the McGill Billboard Corpus \cite{burgoyne_expert_2011} and CoCoPops \cite{arthur_coordinated_2023}. The McGill Billboard dataset has been widely used in numerous MIR tasks, and is known for its comprehensive expert annotations of harmony and form, providing a rich corpus of chord sequences. 
CoCoPops is a meta-corpus of melodic \emph{and} harmonic transcriptions of popular music. It currently consists of two main sub-corpora, the Billboard subset and the Rolling Stone subset. The Billboard subcorpus contains two elements: 1) the entire McGill Billboard Dataset converted into humdrum format, as well as 2) new expert melodic transcriptions of (currently) 214 of these songs.
The Rolling Stone subset contains an additional 200 songs with melody and harmony transcriptions, encoded in the same humdrum format.
We used the Billboard dataset for our single feature (harmony) models, and the CoCoPops dataset in the training and evaluation of our multi-feature models.

\subsection{Experiment Setup}
Our experimental framework is designed to rigorously evaluate and optimize the performance of various models all designed to predict the next chord in a sequence. We employ Optuna, an open-source hyperparameter optimization framework, to systematically search for the optimal model configurations. Each model undergoes a series of trials where Optuna adjusts hyperparameters such as the number of layers, embedding dimensions, and learning rates to maximize prediction accuracy.

For model evaluation, we adopt a k-fold cross-validation approach on the unseen test data, ensuring a robust assessment of each model's accuracy. This method partitions the test data into \(k\) subsets, where each subset serves once as the test set while the remaining \(k-1\) subsets form the training set. The model's accuracy is then averaged over \(k\) runs, providing a comprehensive measure of its generalization capability.

The experiment is orchestrated using \href{https://mlflow.org/}{MLflow}, which tracks each trial's configurations, results, and performance metrics. This integration facilitates a streamlined experiment management process, allowing for efficient comparison between different model architectures and configurations.

The various models, their architecture frameworks, and the code to run the models are all available at the \href{https://github.com/frshdjfry/SeqLab}{SeqLab} repository.

\section{Results}

To test and compare the performance of our proposed models, we conducted extensive experiments across various metrics including accuracy, perplexity, and Word2Vec similarity. \tabref{tab:one_dimension} and \tabref{tab:multi_dimension} encapsulate the comparative results of our models on the McGill Billboard and CoCoPops datasets.

\begin{table}[ht]

\begin{tabular}{lccc}
\hline
\textbf{Model} & \textbf{Acc.} & \textbf{Perp.} & \textbf{W2V Sim.} \\ \hline

Markov Model & 0.140 & 10.31 & 0.860 \\
VO-Markov Model & 0.277 & 2.36 & 0.899 \\
LSTM & 0.191 & 39.76 & 0.879 \\
LSTM + Att. & 0.329 & 10.56 & 0.930 \\
Transformer & 0.321 & 13.81 & 0.930 \\
GPT & 0.301 & 13.24 & 0.925 \\ \hline

\end{tabular}
\caption{Comparative performance of models in one dimension.}
\label{tab:one_dimension}

\end{table}

\begin{table}[ht]

\begin{tabular}{lccc}
\hline
\textbf{Model} & \textbf{Acc.} & \textbf{Perp.} & \textbf{W2V Sim.} \\ \hline

Multi-LSTM & 0.083 & 76.02 & 0.907 \\
Multi-LSTM + Att. & 0.312 & 53.79 & 0.964 \\
Multi-Transformer & 0.271 & 18.65 & 0.956 \\
Multi-GPT & 0.120 & 39.19 & 0.913 \\ \hline

\end{tabular}
\caption{Comparative performance of models in multi-dimension.}
\label{tab:multi_dimension}
\end{table}

\subsection{Model Performances}
The Markov Model and VO-Markov Model demonstrate foundational performance with accuracies of 0.140 and 0.277 respectively, setting a benchmark for more complex models. The LSTM with Attention model shows superior performance on the Billboard dataset with an accuracy of 0.329 and a perplexity of 10.56. Conversely, this model’s performance on CoCoPops is slightly lower with an accuracy of 0.312, suggesting challenges with larger, more complex datasets.

The Transformer and GPT models maintain moderate accuracies of 0.321 and 0.301 across both datasets. The Multi-LSTM with Attention stands out in multi-dimensional analysis on CoCoPops, achieving the highest accuracy of 0.312 and a Word2Vec similarity score of 0.964, confirming its capacity to integrate multiple musical dimensions effectively.

\subsection{Model Interpretation}

Understanding the internal dynamics of predictive models is crucial for improving their accuracy, reliability, and applicability in real-world scenarios. This section explores the model behavior through sensitivity analysis and feature attribution to elucidate how different models process musical sequences and attribute significance to various features.

\subsubsection{Sensitivity Analysis}

To quantitatively assess the influence of each sequence position on the model's prediction accuracy, we used a masking technique where individual positions within a sequence were systematically obscured. We then measured the change in the model's probability of correctly predicting the chord. The influence of each position is calculated as the absolute difference in probabilities, aggregated across all sequences where the model initially made correct predictions:

\[ \text{Influence} = \frac{1}{N} \sum_{i=1}^{N} |\text{Prob}_{\text{original},i} - \text{Prob}_{\text{masked},i}| \]

where \( \text{Prob}_{\text{original},i} \) is the model's probability of correctly predicting the chord for the \(i\)-th sequence without masking, \( \text{Prob}_{\text{masked},i} \) is the probability after masking a position, and \( N \) is the total number of sequences where the original prediction was correct.

\begin{figure*}[ht]
    \centering
    \includegraphics[width=2\columnwidth]{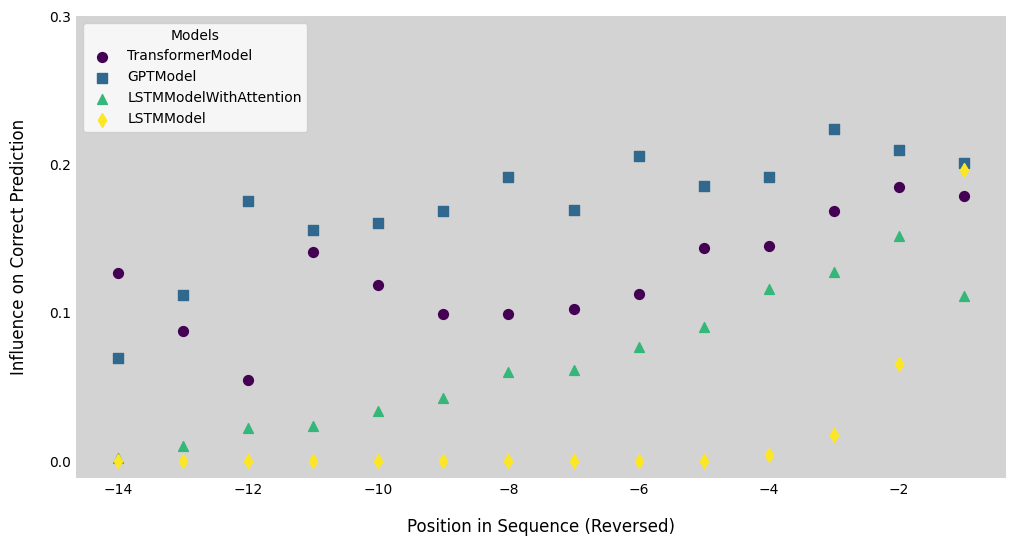}
    \caption{Comparative analysis of position influences on model predictions.}
    \label{fig:comparative_analysis}
\end{figure*}
Figure \ref{fig:comparative_analysis} provides a detailed comparative analysis of how different models utilize positional information within a sequence to predict musical outcomes. The LSTM with Attention and Transformer models, displaying higher overall accuracies, show a pronounced influence from the two preceding positions, underscoring their effective use of recent context for predictions. Notably, these models' influence wanes at earlier positions, highlighting a strategic focus on immediate past information crucial for accuracy.

In contrast, the GPT and basic LSTM models exhibit more erratic influence patterns. The GPT model shows heightened influence at positions -12 to -14, suggesting a struggle to generalize effectively from training data. This might be exacerbated by its primary design for natural language processing, indicating less suitability for nuanced music prediction tasks. The basic LSTM model, lacking positional encoding, demonstrates a scattered influence pattern, confirming its challenges in managing sequence context effectively.

\subsubsection{Feature Attribution}

For models trained on multidimensional data (e.g., chords, melody, rhythm), feature attribution was analyzed through a heatmap representation, focusing on how different features influence the prediction of the next chord.

\begin{figure}
    \centering
    \includegraphics[width=0.9\columnwidth]{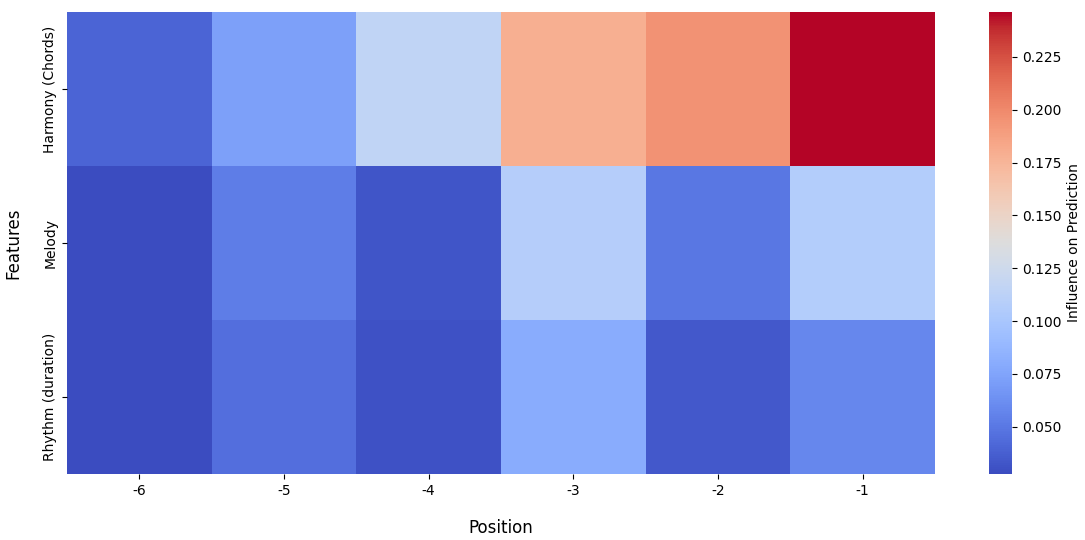}
    \caption{Heatmap showing the influence of each feature position on true prediction accuracy.}
    \label{fig:heatmap}
\end{figure}
Figure \ref{fig:heatmap} illustrates the comparative influence of various musical features on prediction accuracy. Harmony (chords) consistently demonstrates the highest influence across all positions, particularly with the nearest preceding chords where the influence peaks. Melodic elements exhibit moderate influence, with a noticeable impact within the first three preceding positions, highlighting their role in shaping the predictive context. In contrast, rhythmic features, such as note duration, show minimal impact on the prediction outcomes, underscoring their lesser role in determining the immediate musical structure.

\section{Conclusion}

This study has demonstrated the utility and versatility of machine learning models in the domain of music prediction, marking a step forward in bridging the methodological gaps between MIR and music perception.

The differential performance of models across datasets such as McGill Billboard and CoCoPops suggests that the adaptability and effectiveness of these models can vary significantly with the dataset characteristics. While LSTM with Attention models show promise in smaller datasets with their capacity to manage shorter sequence dependencies effectively, Transformer models demonstrate a broader consistency across diverse and larger datasets, suggesting their potential for complex sequence management. 

Furthermore, our employment of sensitivity analysis and feature attribution should be seen as an initial step towards more comprehensive interpretive techniques in the study of ML models in music cognition. These methods, while illustrative in this context, require further development to provide deeper and more conclusive insights into how different features influence model predictions.

In conclusion, the findings from this study encourage cautious optimism about integrating ML models into music cognition research. They highlight the potential of these models to contribute to our understanding of musical structure and perception but also underscore the need for continued refinement and critical evaluation.

\newpage

\bibliography{striking}

\end{document}